\documentstyle[11pt,newpasp,twoside,epsf]{article}
\def\edcomment#1{\iffalse\marginpar{\raggedright\sl#1\/}\else\relax\fi}
\reversemarginpar

\begin{document}
\title{Old White Dwarfs as a Microlensing Population}
 \author{Brad M. S. Hansen}
\affil{Department of Astrophysical Sciences, Peyton Hall,
 Princeton University, 
Princeton, NJ, 08544, USA}

\begin{abstract}
A popular interpretation of recent microlensing studies of the line
of sight  towards the Large Magellanic Cloud invokes a population of
old white dwarf stars in the Galactic halo. Below I review the basic
properties of old white dwarf stars and the ongoing efforts to detect
this population directly.
\end{abstract}

\section{Introduction}

Other authors in this volume cover the microlensing motivations much
better than I can, so I shall suffice to remind you that one possible
explanation of the microlensing events towards the LMC invokes a population
of objects in the range $0.3-0.8$M$_{\odot}$. Potentially these could be
either normal hydrogen-burning stars or white dwarfs, the burnt-out remnants
of stellar evolution. To distinguish these populations, we turn to direct
searches at optical wavelengths, since the latter population is 
$10^{-3}-10^{-4}$ as bright as the former. The number counts of faint
red stars suggest that hydrogen burning stars cannot account for the
microlensing population (Bahcall et al 1994; Graff \& Freese 1996).
 The question I wish to
address is how well one can constrain the white dwarf hypothesis by 
similar means.

\section{White Dwarf Cooling}

To derive a constraint from direct optical searches, we need to know how to
recognise white dwarfs. White dwarfs of moderate age ($<10$~Gyr) emit
approximately as black bodies. However, once the white dwarf cools to
effective temperatures below $\sim 5000$~K, the atmospheric hydrogen
resides in molecular form. This has dramatic consequences for the appearance
of the white dwarf (Hansen 1998, 1999a; Saumon \& Jacobsen 1999), because the
dominant opacity source in such an atmosphere is collisionally induced
absorption by H$_2$, which absorbs primarily in the near infra-red. 
For black bodies of  effective temperatures $\sim 3000-5000$~K the peak
of the spectrum lies in the same wavelength region, so that the increased
absorption leads to dramatic deviations from the traditional assumption
of black body colours. The general trend of the colour evolution is that
the flux is forced to emerge preferentially blueward of $\sim 0.8 \mu$m.
With the evolution of the black body flux redward, the peak of the spectrum
for old white dwarfs lies around $0.6 \mu$m.

A correct quantitative analysis of the atmospheric conditions is not only
important for determining the colours of the white dwarf, but is also of
critical importance for the cooling evolution. For white dwarfs cooler than
$T_{\rm eff} \sim 6000$~K, the internal structure has no radiative zones. The
bulk of the white dwarf mass resides in a degenerate core which is approximately
isothermal due to the efficient thermal conduction by the electrons near the
surface of the Fermi sea. This isothermal core is joined to a thin convective
envelope which extends all the way to the photosphere. Thus, the entire white
dwarf structure is critically dependant on the surface boundary condition and
hence the atmospheric conditions. It must be noted that, while sophisticated
analyses of white dwarf atmospheres have been around for several years
(e.g. Bergeron, Saumon \& Wesemael 1995), the white dwarf cooling calculations
have lagged significantly behind, employing grey atmosphere boundary conditions.
Thus, while one may trust the masses and gravities inferred in recent atmospheric
analyses, one cannot completely trust the ages inferred for the older white dwarfs 
in most recent papers. To my knowledge, the only cooling calculations using
boundary conditions based on proper atmosphere models are those of Hansen (1999a)
and Salaris et al (2000). For isochrones and luminosity functions
using the Hansen (1999a) models, see Richer et al (2000).

The models described above concern white dwarfs with pure hydrogen atmospheres.
Empirically, a significant fraction of known white dwarfs appear to have no
hydrogen in their atmospheres (Liebert et al 1979; Bergeron, Ruiz \& Leggett 1997).
When pure helium atmospheres reach effective temperatures $<5000$~K, they have
no molecular component to provide significant opacity and, as a result, the
photosphere lies at much higher densities, where the beginnings of pressure
ionization start to provide a small ionized component. This higher density and
the concomitant difference in the boundary condition results in much faster
cooling for old white dwarfs with helium atmospheres. Furthermore, these
objects are expected to appear approximately as black bodies (contrary to the
hydrogen atmosphere case) but will be hard to detect because they cool much
faster.

The matter of a component of the white dwarf population sporting helium
atmospheres is one that is often not mentioned in the recent studies attempting
to tie direct searches to microlensing populations. It is important to note
that this represents an essentially unobservable (due to their much more rapid
cooling) component and, as such, will always introduce some uncertainty into
the comparison.

\section{Observational Progress}

\subsection{The story thus far}

\begin{figure}
\plotone{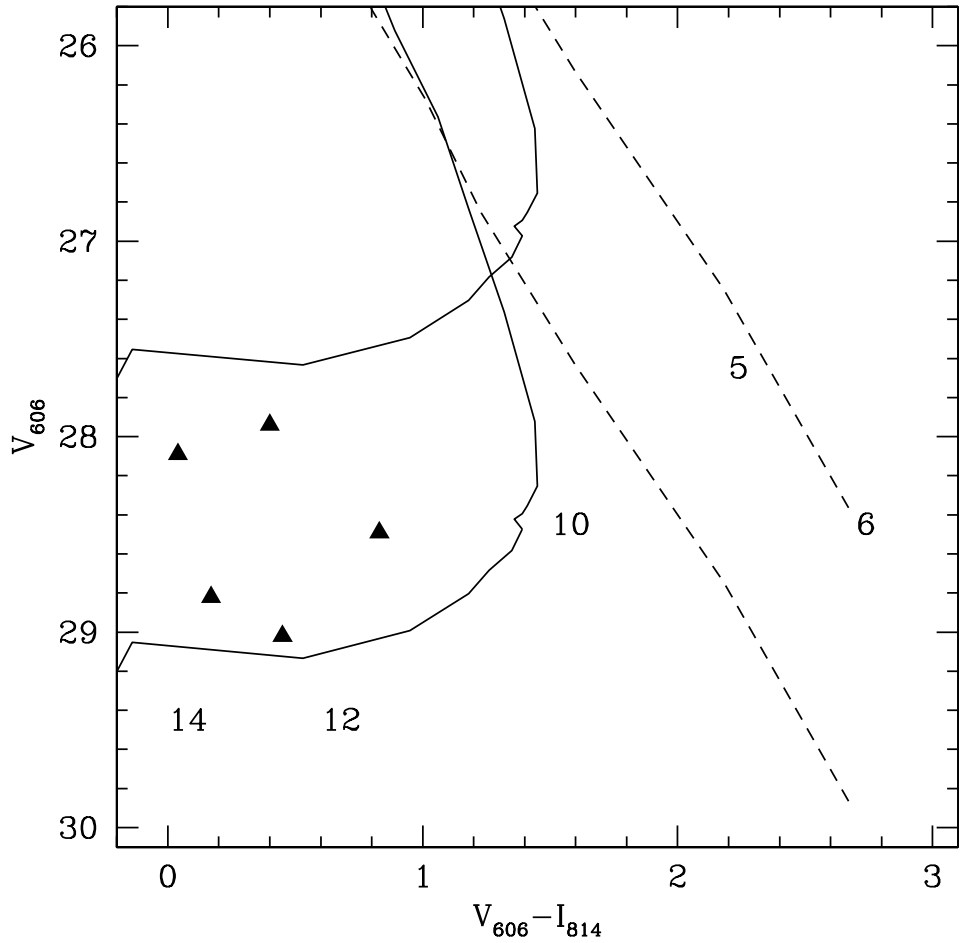}
\caption{The source of all the fuss. The five points are the magnitudes \& colours
of the 5 proper motion objects of Ibata et al (1999). The solid curves describe the
cooling of a 0.6 solar mass white dwarf with hydrogen atmosphere located at 1 and 2
kpc. Representative ages are indicated along the lower curve. The dashed curves
are the same but for a pure helium atmosphere. The much more rapid cooling is
evident.
}
\end{figure}      

The last couple of years have been fun for those interested in the question
of halo white dwarfs. The realisation that old white dwarfs would appear
somewhat bluer than black bodies meant that previous constraints on the local
number density based on Hubble Deep Field point source counts were somewhat overstated.
Much excitement was generated by the detection of proper motions of several
faint blue point sources in the HDF North (Ibata et al 1999). The colours and
magnitudes of these objects are consistent with old white dwarfs $\sim 12$~Gyr
old at distances $\sim 1$~kpc. Given the radical consequences of this observation,
some kind of confirmation would be nice. A third epoch, to confirm the measurement,
has been approved but was postponed due to the HST gyro problems in December 1999.
Nevertheless, the veracity of the HDF proper motions should be addressed within
the next year or so.

In the interim, several other interesting results have come to light. Most
importantly, Hodgkin et al (2000) presented a near-infrared spectrum of a nearby
cool white dwarf, demonstrating the reality of the flux-suppression due to 
molecular hydrogen absorption. This provides a welcome check that the 
theory behind the predicted colour change is at least qualitatively correct.
The high proper motion of this object also suggests membership of the Galactic halo.
Another object showing such a long wavelength depression is LHS~3250 (Harris et al
1999)

A further interesting indirect argument supporting the white dwarf hypothesis
comes from Mendez \& Minniti (2000), who assert that the number counts of faint blue
point sources (the class amongst which Ibata et al discovered proper motions)
in the Hubble Deep Field South is approximately twice that in the HDF North. This
is consistent with the objects (whatever they are) being Galactic in origin, since
the southern field looks in a direction closer to the Galactic centre and consequently
the stellar density is expected to be higher. 

A note of caution, however, is sounded by Flynn et al (2000), who find little evidence
for a population equivalent to that of Ibata et al in the Luyten proper motion 
survey. They conclude that hydrogen atmosphere white dwarfs cannot comprise a
significant fraction of the Galactic halo. On the other hand, a new search of
wide field photographic plates by Ibata et al (2000) has uncovered several
high proper motion objects, at least two of which are spectroscopically confirmed
to be white dwarfs and with some modicum of IR flux suppression. 

In my not-completely-unbiased interpretation of the above, the evidence in
favour of the white dwarf hypothesis is encouraging, although not conclusive.
The recent results of Hodgkin et al and Ibata et al (2000) have
demonstrated at least the existence of a population of high proper motion
(and thus probably halo) white dwarfs. The question is now a matter of number
density. It is on this point that we must await further data. 

\subsection{Future prospects}

While the above results indicate that some white dwarfs do show
halo kinematics, the fact 
 that there is little evidence for a large halo white
dwarf population in existing large-scale proper motion
surveys casts some doubt that this population may be responsible
for the microlensing results. This last statement relies heavily
on the data set of Luyten (1979), so it would be nice to check this
with another large scale survey.

Towards this end, a program led by Peter Stetson has begun at CFHT
to image almost 20 square degrees to $V \sim 25$. Repeated with
an interval of 2-3 years, this program should provide a conclusive
test of the white dwarf halo hypothesis. Using the new MACHO mass
fraction estimate ($\sim 20\%$) and assuming only half of all white
dwarfs have hydrogen atmospheres, this program is still expected to
find about 8 white dwarfs per square degree. 

\begin{figure}
\plottwo{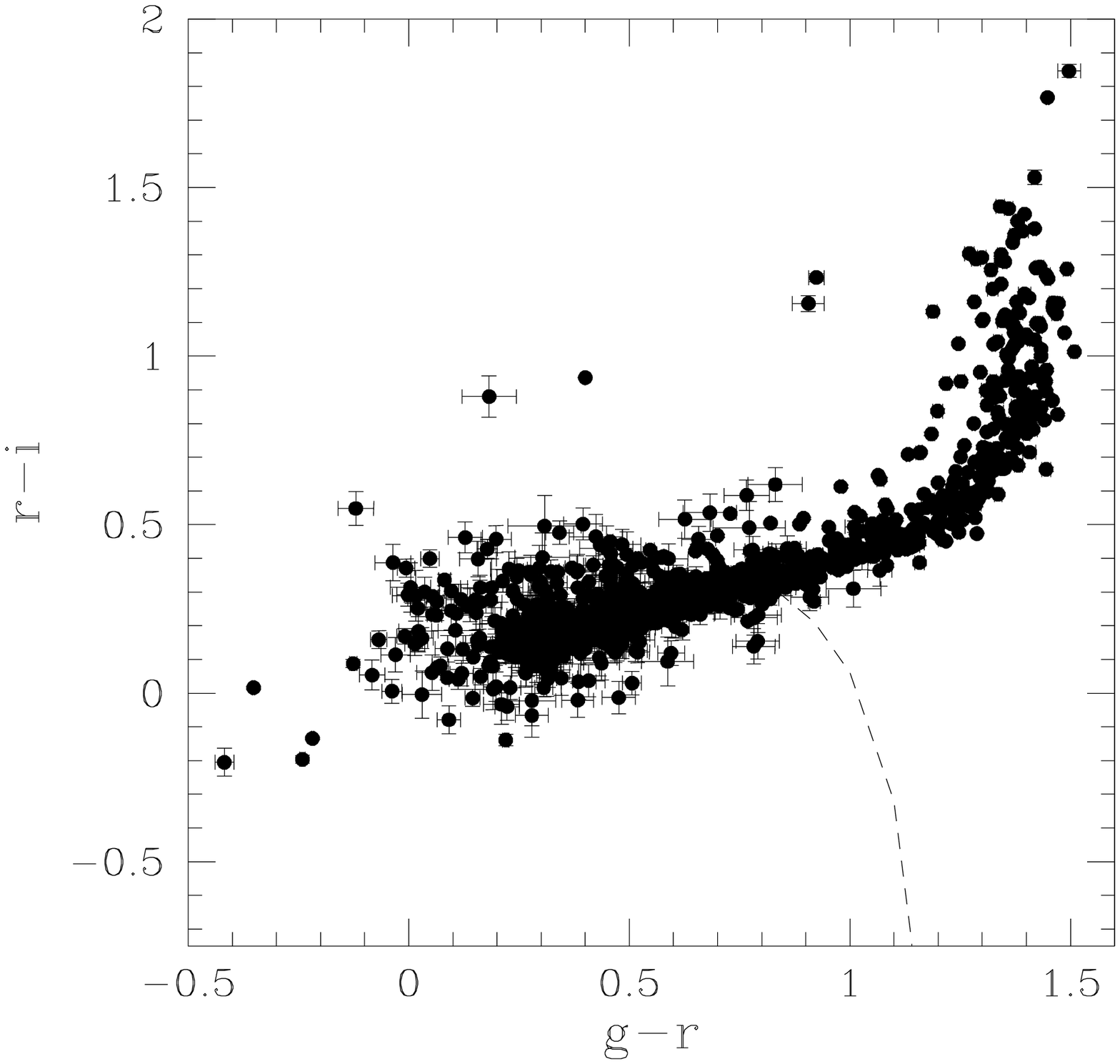}{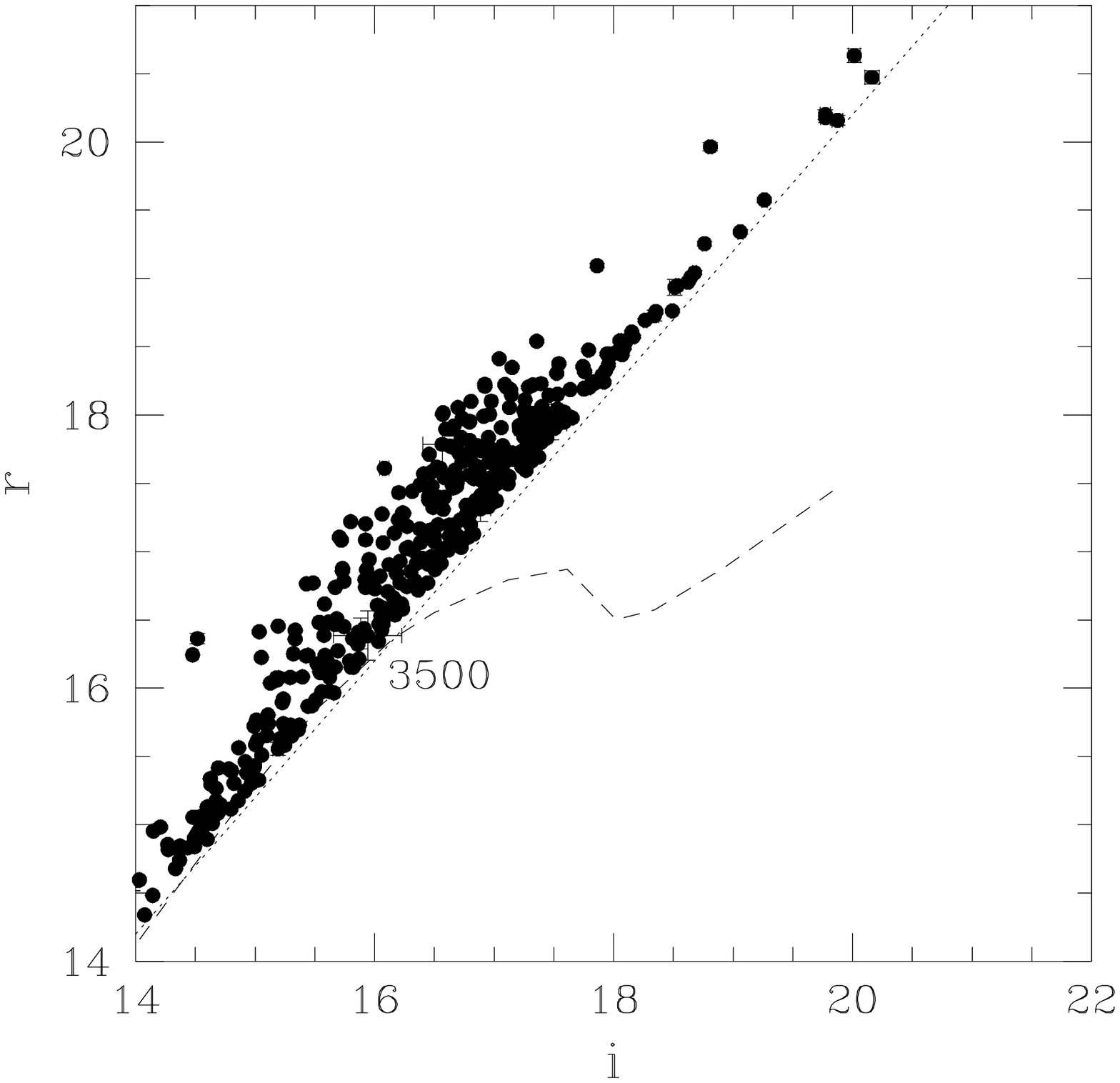}
\caption{The left hand panel shows the SDSS stellar locus 
(York et al 2000) in the
g-r/r-i plane, with the evolution of a representative white dwarf
shown by the dashed line. Thus, it is possible to select very cool
white dwarfs by colour alone. However, objects that cool are also
faint. The right hand panel shows the apparent magnitudes of the 
stars with g-r$>$0.8 as well as the dashed line for the cooling of a 0.6~$M_{\odot}$
white dwarf at 10~pc (i.e. the absolute magnitude). The dotted line shows
the apparent magnitude of a 3500~K white dwarf at various distances.}
\end{figure}

Another fascinating prospect is the hope of finding such old white
dwarfs in the Sloan Digital Sky Survey (SDSS). Although this comprehensive
large scale survey will not provide proper motions as a matter of course,
some proper motion information is potentially available from comparison
with Palomar Sky Survey plates. Furthermore, there will be a strip of sky
in the southern hemisphere that will be multiply imaged. Even without
proper motions, very old white dwarfs are potentially detectable by
colours alone in the SDSS five-band photometric system. Photometric
selection alone is unlikely to provide a complete census as only the
oldest white dwarfs will stand out from the locus of other stars
(the problem is that even the reddest bandpass doesn't extend significantly
beyond $1\mu$m). Nevertheless, very old white dwarfs are interesting
in their own right.

Figure~2 shows a representative stellar locus and how one
might select old white dwarfs based on colours. The colour selection 
suggests that only white dwarfs cooler than 3500~K can be selected in
this manner. Using an i-band detection limit of 18, we find such objects
can be detected to distances $\sim 30$~pc. If the entire MACHO fraction
(taken to be 20\%) were in 0.5$M_{\odot}$ white dwarfs, 50\% of whom
had hydrogen atmospheres, and all had 3500~K temperatures, we would expect
approximately 43 white dwarfs in the Sloan survey, or 1 every
480 square degrees. This is probably an overestimate. Hotter white dwarfs
cannot be distinguished from the stellar locus (it is worth noting that
the Hodgkin et al and Ibata et al detections all correspond to 
$T_{\rm eff} \sim 3500-4000$~K, so in the marginal region)
 and cooler objects are fainter and thus the
effective volume is smaller. Nevertheless, this simple estimate offers
encouragement that at least some white dwarfs are potentially detectable
in this manner. More detailed calculations are underway to provide a more
robust estimate.

Another large scale proper motion survey that should offer interesting
constraints is that being undertaken by the EROS project (Goldman 1998).
With a survey area of 350 square degrees and an I-limit of 20.5, we can
use the same approximate scenario as above to estimate $\sim 23$ white
dwarfs in their sample.
 This is a somewhat more realistic ballpark number in this case as the
proper motion selection will allow detection of hotter objects. 

The bottom line of the above is that, although uncertainty about mass distributions
and chemical compositions make prognostication difficult, the fact that naive
estimates lead to predictions of 10-100 detections in ongoing surveys suggests
that we should see something if the hypothesis of a significant white dwarf
halo is correct.

\section{Beige Dwarfs}

Although it was not my intention to address this subject at the conference,
it was raised a couple of times, so I will conclude with a few remarks about
``Beige'' dwarfs (Hansen 1999b). Given that brown dwarfs are strongly ruled out
by the microlensing timescales, red dwarfs by direct observations and white
dwarfs (as well as neutron stars)
 are an uncomfortable fit due to issues of chemical pollution, it appears
that there are no surviving baryonic candidates for halo MACHOs. However, there
is one remaining possibility. Lenzuni, Chernoff \& Salpeter (1992) demonstrated
that one could circumvent the traditional hydrogen burning limit by starting
with a brown dwarf and accreting material slowly enough that it could cool.
In this fashion one could construct a degenerate hydrogen/helium object with
mass $>0.1 M_{\odot}$. If one could build such objects to masses $>0.3 M_{\odot}$,
they would make ideal MACHO candidates as they would be faint and would have
no nuclear burning history and thus no chemical pollution problem. The term
``Beige'' dwarf comes from the superposition of brown dwarf and white dwarf 
characteristics.
 The primary
problem for this scenario is that there is little evidence that this mode of
``star'' formation was ever important. 

The existence of Beige dwarfs is observationally testable. They have similar
radii to white dwarfs but can only exist at effective temperatures
$T_{\rm eff} < 2000$~K, so the detection of hotter objects would suggest white
dwarfs. Indeed, the spectroscopically confirmed objects of Hambly et al
and Ibata et al (2000) indicate temperatures $\sim 3500-4000$~K. Therefore
these are unlikely to be Beige dwarfs. 

\acknowledgements 
I would like to thank the conference organisers for a thoroughly enjoyable
(and efficiently run) meeting.
Support for this work was provided by NASA through Hubble Fellowship grant 
\#HF-01120.01-99A,
from the Space Telescope Science Institute, which is operated by the
 Association of Universities
for Research in Astronomy, Inc., under NASA contract NAS5-26555.



\begin{references}
\reference{Bahcall, J. N., Flynn, C., Gould, A. \& Kirhakos, S. 1994, ApJ, 435, L51}
\reference{Bergeron, P., Ruiz, M.-T. \& Leggett, S. K. 1997, ApJS, 108, 339}
\reference{Bergeron, P., Saumon, D. \& Wesemael, F. 1995, ApJ, 443, 764}
\reference{Flynn, C., Sommer-Larsen, J., Fuchs, B., Graff, D. S. \& Salim, S. 2000,
astro-ph/9912264}
\reference{Goldman, B., in The Third Stromlo Symposium: The Galactic Halo, eds.
Gibson, B. K., Axelrod, T. S. \& Putman, M. E., ASP Conference Series Vol 165,
p 413; San Francisco}
\reference{ Graff, D.S. \& Freese, K. 1996, ApJ, 435, L516}
\reference{ Hansen, B. M. S. 1998, Nature, 394, 860}
\reference{ Hansen, B. M. S., 1999a, ApJ, 520, 680}
\reference{ Hansen, B. M. S., 1999b, ApJ, 517, L39}
\reference{ Harris, H. C., Dahn, C. C., Vrba, F. J., Henden, A. A., Liebert, J.,
Schmidt, G. D. \& Reid, I. N. 1999, ApJ, 524, 1000}
\reference{ Hodgkin, S. T., Oppenheimer, B. R., Hambly, N. C., Jameson, R. F.,
Smartt, S. J. \& Steele, I. A. 2000, Nature, 403, 57}
\reference{ Ibata, R. A., Richer, H. B., Gilliland, R. L. \& Scott, D. 1999, ApJ, 524, L95}
\reference{ Ibata, R. A., Irwin, M., Bienayme, O., Scholz, R. \& Guibert, J. 2000,
ApJ, 532, L41}
\reference{Lenzuni, P., Chernoff, D. F. \& Salpeter, E. E., 1992, ApJ, 393, 232}
\reference{ Liebert, J., Dahn, C. C., Gresham, M. \& Strittmatter, P. A. 1979,
ApJ, 233, 226}
\reference{Luyten, W. J. 1979, {\em LHS Catalogue: A catalogue of stars with proper motions
exceeding 0.5" annually}, University of Minnesota, Minneapolis}
\reference{Mendez, R. A. \& Minniti, D. 2000, ApJ, 529, 911}
\reference{ Richer, H. B., Hansen, B., Limongi, M., Chieffi, A., Straniero, O.
\& Fahlman, G. G. 2000, ApJ, 529, 318}
\reference{Salaris, M., Garcia-Berro, E., Hernanz, M., Isern, J. \& Saumon, D. 2000, submitted to ApJ}
\reference{Saumon, D. \& Jacobsen, S. B. 1999, ApJ 511, L107}
\reference{York, D. et al 2000, AJ submitted}
\end{references}
\end{document}